\documentstyle[twoside,fleqn,espcrc2,epsf,rotate,times]{article}
\newcommand{\lt}{\left}
\newcommand{\rt}{\right}
\newcommand{\ov}{\overline}

\newcommand{\eq}[1]{(\ref{#1})}

\def\openone{\leavevmode\hbox{\small1\kern-3.8pt\normalsize1}}%
\newcommand{\nn}{\nonumber \\}
\newcommand{\no}{\nonumber }

\newcommand{\imag}{{\rm Im}\,}
\newcommand{\real}{{\rm Re}\,}

\newcommand{\epm}[2]{
 \raisebox{-0.5ex}{\shortstack[l]{$\scriptstyle+#1$\\$\scriptstyle-#2$}}
                    }

\newcommand{\bra}[1]{\langle \, #1 \, | }
\newcommand{\ket}[1]{| \, #1 \, \rangle }

\newcommand{\e}{\epsilon}
\newcommand{\gev}{\,\mbox{GeV}}
\newcommand{\mev}{\,\mbox{MeV}}

\newcommand{\prl}{Phys.\ Rev. Lett.}
\newcommand{\prd}{Phys.\ Rev.~D}
\newcommand{\plb}{Phys.\ Lett.~B}
\newcommand{\npb}{Nucl.\ Phys.~B}

\title{
~\vspace{-2.2truecm}\\
{\normalsize 
Fermilab-Conf-99/288-T \hfill hep-ph/9910257
\hspace{2truecm}~}\\[1.5truecm]
Theoretical status of $\epsilon^\prime/\epsilon$
        \thanks{Invited plenary talk at {\it QCD 99, 
		July 7-13th 1999, Montpellier, France}
}}

\author{Ulrich Nierste\\
	Fermilab Theory Division, MS106, Batavia IL60510, USA}

\begin{document}

\begin{abstract}
I give a detailed introduction into the theoretical formalism for
$\epsilon^\prime/\epsilon$, which measures direct CP-violation in $K
\rightarrow \pi\pi$ decays. The current status of hadronic matrix
elements and the strange quark mass is discussed.  Several possible
explanations of the unexpectedly high experimental results for
$\epsilon^\prime/\epsilon$ are pointed out: A small strange quark
mass, an enhancement of the hadronic parameter $B_6^{(1/2)}$ from the
$\sigma$ resonance, an underestimate of isospin breaking and possible
new physics contributions in the $\ov{s}dZ$-vertex and the
$\ov{s}d$-gluon-vertex.
\end{abstract}

\maketitle

\section{Setting the scene}
\addtolength{\textheight}{-53pt}
\addtolength{\topmargin}{53pt}
This year has surprised us with new experimental results for
$\epsilon^\prime/\epsilon$, which measures direct CP-violation in $K
\rightarrow \pi\pi$ decays. Until February 1999 the experimental
situation was inconclusive: While the CERN NA31 \cite{na} experiment
has clearly shown a non-vanishing $\epsilon^\prime/\epsilon$, the
Fermilab E731 result \cite{e731} was consistent with zero. With the
new measurements of the Fermilab KTeV \cite{ktev} and the CERN NA48
\cite{na48} collaborations this issue is settled now in favour of the
non-zero NA31 result. The new world average reads
\begin{eqnarray} 
\; \real  \frac{\epsilon^\prime}{\epsilon} &=&  \lt(21.1 \pm 4.6\rt)
\cdot 10^{-4} . \label{exp}
\end{eqnarray} 
The establishment of direct CP-violation rules out old superweak
models. Yet while the Standard Model predicts a non-vanishing
$\epsilon^\prime/\epsilon$, the value in \eq{exp} came as a surprise
to many theorists, as it exceeds most theoretical predictions.  In
the following I will try to illuminate the possible sources of this
discrepancy.
   
\begin{figure}
\centerline{\epsfxsize=0.23\textwidth
\rotate[r]{\epsffile{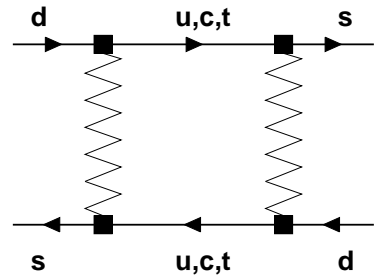}}}
\vspace{-9pt} 
\caption{Standard Model $\Delta S=2$ box diagram. The zigzag lines denote 
	W-bosons.}\label{box}
\end{figure}

\begin{figure} 
\centerline{\epsffile{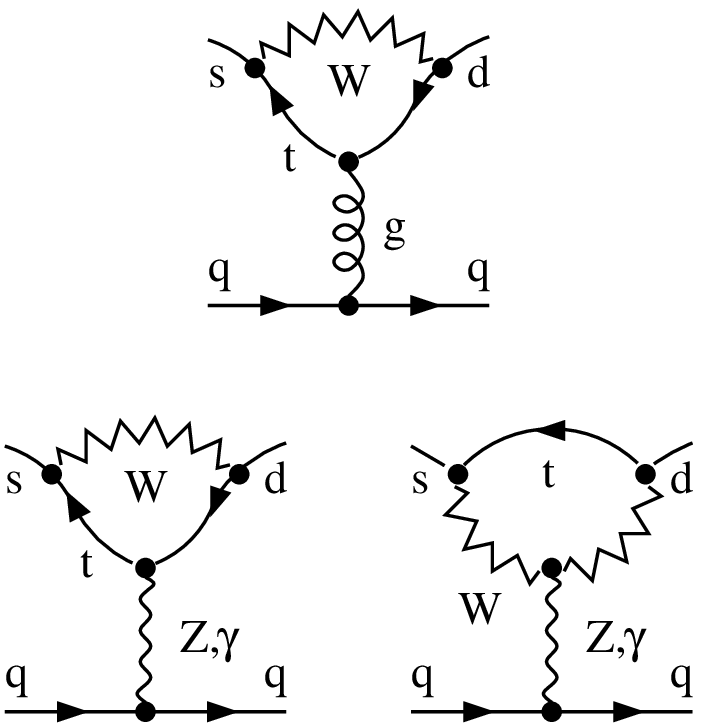}}
\vspace{-9pt} 
\caption{The $\Delta S=1$ diagrams dominating $\e^\prime$ in the
Standard Model: QCD and electroweak penguin diagrams.}\label{peng}
\end{figure}

The flavour eigenstates of neutral K mesons,
\begin{eqnarray}
\qquad \ket{K^0} \sim \ov{s}d &\quad \mbox{and}\quad & 
\ket{\ov{K}{}^0} \sim s\ov{d}, \no
\end{eqnarray} 
combine into mass eigenstates  $\ket{K_L}$ and $\ket{K_S}$ due to  
$K^0$--$\ov{K}{}^0$ mixing depicted in Fig.~\ref{box}. 
CP-violation in neutral Kaon decay is illustrated in Fig.~\ref{figcp}.
$\ket{K_L}$ and $\ket{K_S}$ are not CP eigenstates, because the
$\Delta S=2$ amplitude inducing $K^0$--$\ov{K}{}^0$ mixing violates
CP. This {\it indirect CP-violation}\ has been discovered in 1964 by
Christenson, Cronin, Fitch and Turlay \cite{ccft}. By contrast {\it
direct CP-violation}\ denotes CP-violation in the $\Delta S=1$
amplitude triggering the decay. It allows a CP-odd initial state to
decay into a CP-even final state and vice versa. The dominant
contributions to direct CP-violation in the Standard Model are
depicted in Fig.~\ref{peng}.

\begin{figure*} 
\begin{displaymath}
\qquad\qquad\qquad
\begin{array}{rcc@{}cc}
\ket{K_L} & \propto 
& \left[    
   \underbrace{   \ket{K^0} - \ket{\ov{K^0}} } +\right. & \ov{\epsilon} 
  \left.    \left( \underbrace{ \ket{K^0} + \ket{\ov{K^0}} }  \right)  
   \right] &   
  \\[5mm] 
&&\!\!\! \overbrace{\mbox{CP=--1 }} &\, \overbrace{\mbox{CP=+1 }}& 
\nonumber \\ 
\multicolumn{2}{c}{\mbox{Decay:}} & \multicolumn{2}{c}{
\parbox[t]{60mm}{
\setlength{\unitlength}{1mm} 
\begin{picture}(60,10)  
\put(12,0){\line(0,1){10}}
\put(12,0){\vector(1,0){50}}  
\put(44,3){\line(0,1){7}}
\put(44,3){\vector(1,0){18}}
\put(47,5){\mbox{indirect}}  
\put(15,2){\mbox{direct}}  
\end{picture}}} & \mbox{\large$\pi^0 \pi^0$} 
\\
&&&&\mbox{(CP=+1)} 
\end{array} 
\end{displaymath}
\caption{Direct and indirect CP-violation in $K_L$ decay. The pattern
for $K_S$ decay is analogous. }\label{figcp}
\end{figure*} 

In order to disentangle and quantify the two types of CP-violation one
introduces isospin amplitudes:
\begin{eqnarray}
\; {\cal A}  \lt( K^0 \rightarrow \pi^0 \pi^0 \rt) &=& 
 \sqrt{\frac{2}{3}} A_0 e^{i \delta_0} -  
 \sqrt{\frac{4}{3}} A_2 e^{i \delta_2} \nn
\; {\cal A}  \lt( K^0 \rightarrow \pi^+ \pi^- \rt) &=& 
 \sqrt{\frac{2}{3}} A_0 e^{i \delta_0} +   
 \sqrt{\frac{1}{3}} A_2 e^{i \delta_2} \nn
\; {\cal A}  \lt( K^+ \rightarrow \pi^+ \pi^0 \rt) &=& 
 \sqrt{\frac{3}{2}} A_2 e^{i \delta_2} .\label{iso}
\end{eqnarray}
In the limit of exact isospin symmetry, the hadronic final state
interaction only produces non-vanishing strong phases $\delta_0$ and
$\delta_2$ of the isospin $I=0$ and $I=2$ amplitudes in \eq{iso}.
Rescattering between the $\lt(\pi \pi\rt)_{I=0}$ and $\lt(\pi
\pi\rt)_{I=2}$ state stems only from small isospin breaking effects,
which will appear as a correction term in the formula for
$\epsilon^\prime$. The isospin amplitudes $A_0$ and $A_2$ are still
complex, they contain the information on the CP-violating weak
phases. For the CP-conjugated amplitudes to \eq{iso} describing
$\ov{K}{}^0$ and $K^-$ decay one needs to replace $A_0$ and $A_2$ in
\eq{iso} by their complex conjugates without changing the strong
phases $\delta_0$ and $\delta_2$. In phenomenological analyses of
$\epsilon^\prime$ one takes the real parts of $A_0$ and $A_2$ from
experiment:
\begin{eqnarray}
\; \real A_0  &=& 3.33\cdot 10^{-7} \gev \nn
\; \omega &=& \frac{ \real A_2 }{ \real A_0} \; = \; \frac{1}{22.2} .
\label{d1/2}
\end{eqnarray} 
The smallness of $\omega$ has motivated the phrase ``$\Delta I=1/2$
rule'', because isospin changes by a half unit in the dominant 
$K\rightarrow \lt(\pi \pi\rt)_{I=0}$ decay. 
 
Direct CP-violation requires the interference of at least two amplitudes
with different strong (and weak) phases. Hence 
\begin{eqnarray}
\; \e 
&=& 
 \frac{ {\cal A} \lt( K_L \rightarrow (\pi \pi)_{I=0} \rt)  }{
         {\cal A} \lt( K_S \rightarrow (\pi \pi)_{I=0} \rt)   } 
\nn
&=&  \lt( 2.280\pm 0.013 \rt) \cdot 10^{-3} \, e^{i\pi/4}
\label{expeps}
\end{eqnarray}
purely measures indirect CP-violation. Direct CP-violation is encoded 
in $\e^\prime$ with 
\begin{eqnarray}
\frac{\e^\prime}{\e} &=& \frac{1}{\sqrt{2}} \, \lt[
    \frac{ {\cal A} \lt( K_L \rightarrow (\pi \pi)_{I=2} \rt)  }{
         {\cal A} \lt( K_L \rightarrow (\pi \pi)_{I=0} \rt)   } 
	\rt.\nn 
&& \phantom{ \frac{1}{\sqrt{2}} }
  \lt.  - \frac{ {\cal A} \lt( K_S \rightarrow (\pi \pi)_{I=2} \rt)  }{
         {\cal A} \lt( K_S \rightarrow (\pi \pi)_{I=0} \rt)   } \rt] 
	\nn
&=&  \frac{1}{\sqrt{2} \lt| \e \rt|} \, \imag \lt( \frac{A_2}{A_0}\rt) \nn
&=& \frac{\omega}{\sqrt{2} \lt| \e \rt|} 
		\lt[ \frac{\imag A_2}{\real A_2} -  
		     \frac{\imag A_0}{\real A_0} \rt] .\label{ep}
\end{eqnarray} 
The phase of $\e^\prime$ is $\pi/2+\delta_2-\delta_0$ and numerically
coincides with the phase of $\e$.  

\begin{figure} 
\centerline{\epsfxsize=0.4\textwidth
		\epsffile{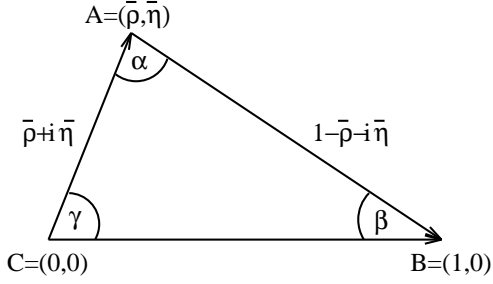}}
\caption{The unitarity triangle. CP-violating observables are usually
expressed in terms of its height $\ov{\eta}$ or one of its angles.
The lengths of its sides can be determined form CP-conserving
quantities. }\label{ut}
\end{figure}

The Standard Model predicts both indirect and direct CP-violation. The
necessary CP-violating phases originate from the elements $V_{u_id_j}$
of the Cabibbo-Kobayashi-Maskawa (CKM) matrix. They enter the diagrams
in Figs.~\ref{box} and \ref{peng} at the couplings of the W-boson to
the quarks. The single complex phase $\gamma$ in the CKM matrix must
fit all CP-violating observables in $K$- and $B$-physics and moreover
is also constrained by CP-conserving quantities.  This will be a
litmus test for the standard model and eventually open the door to new
physics, once the dedicated B-physics experiments at $e^+e^-$
\cite{babar} and hadron colliders \cite{fnal} deliver data and the
rare Kaon decays $K^+ \rightarrow \pi^+ \ov{\nu} \nu$ \cite{k+} and
$K_L \rightarrow \pi^0 \ov{\nu} \nu$ \cite{k0} are precisely measured.  
The corresponding phenomenology is condensed into the
unitarity triangle depicted in Fig.~\ref{ut}. Its apex
$(\ov{\rho},\ov{\eta})$ is defined by
\begin{eqnarray}
\;
\ov{\rho}+ i \ov{\eta} \!\!&=&\!\! - \frac{V_{ud} V_{ub}^*}{V_{cd} V_{cb}^*}, 
\quad \mbox{and} \quad  \gamma\;=\; 
\arctan \lt( \frac{\ov{\eta}}{\ov{\rho}} \rt) . \no
\end{eqnarray}
In addition to $\ov{\rho}$ and $\ov{\eta}$ one needs two more
quantities to parameterize the CKM matrix, $\lambda\simeq
|V_{us}|=0.22$ and $A=|V_{cb}|/\lambda^2=0.80\pm0.05$. The
Wolfenstein parameterization \cite{wolf} is an expansion of the CKM
matrix in terms of $\lambda$ to the third order. In CP studies higher
orders in $\lambda$ must be included \cite{blo}. 
In Kaon physics CP-violating quantities are conveniently expressed in
terms of 
\begin{eqnarray}
\; \imag \lambda_t &=& - \imag \lambda_c, \quad \mbox{where} \quad
	\lambda_{i} \; = \; V_{is}^* V_{id}. \label{iml}
\end{eqnarray}
$\lambda_t$ is the CKM factor of the penguin diagrams in
Fig.~\ref{peng} and $\epsilon^\prime$ is proportional to 
$\imag \lambda_t$. Useful relations are 
\begin{eqnarray}
\; \imag \lambda_t \!\!\!\!&\simeq & \!\!\!\!|V_{cb}| |V_{ub}| \sin \gamma 
       \; \simeq \; A^2 \lambda^5 \ov{\eta} \; \simeq \; J_{CP}/\lambda
\label{imlat}
\end{eqnarray}
Here ``$\simeq$'' means equal up to corrections of order 
$\lambda^2$ or smaller. $J_{CP}$ is the Jarlskog invariant of
CP-violation. $\ov{\rho}$ and $\ov{\eta}$ are best suited for the
phenomenology of CP-violation in B-physics, yet in Kaon physics
CP-violation stems form loop-induced $s\rightarrow d$ transitions and 
$\imag \lambda_t$ is the natural parameter here. Using $\ov{\eta}$ 
instead artificially introduces high powers of $\lambda$ and $A$ and
the associated uncertainties into the problem, see \eq{imlat}. We will also 
encounter 
\begin{eqnarray}
\; \real \lambda_t &\simeq& - \lt(1-\ov{\rho}\rt) A^2 \lambda^5,
\qquad \real \lambda_c \; \simeq \; -\lambda . \label{relat}
\end{eqnarray}

Next we cast \eq{ep} into the form 
\begin{eqnarray}
\; \frac{\varepsilon '}{\varepsilon} &=& 
\imag \lambda_t  \:
\frac{G_F \omega}{2\, |\epsilon|\, \real A_0 } \:
 \left[ \Pi_0 - \frac{1}{\omega} \: \Pi_2 \right] \; .
\label{eps'2}
\end{eqnarray}
Here $G_F$ is the Fermi constant.  
$A_0$ and $A_2$ are contained in 
\begin{eqnarray}
\; \Pi_0 &=&  \sum_{i=3}^{10} y_i \langle Q_i \rangle_0 
	\lt( 1- \Omega \rt) \nn
& \approx & y_6 \langle Q_6 \rangle_0 
        \lt( 1- \Omega \rt) \nn 
\; \Pi_2 &=& \sum_{i=7}^{10} y_i  \langle Q_i \rangle_2 
	\; \approx \; y_8   \langle Q_8 \rangle_2 .
\label{p0}
\end{eqnarray}
In \eq{p0} an operator product expansion has been performed to
separate short and long distance physics.  The short distance physics
is contained in the Wilson coefficients $y_i$. The heavy masses of the
top quark and the W-boson propagating in the loop diagrams of
Fig.~\ref{peng} enter these coefficients.  The $y_i$ can be reliably
calculated in perturbation theory, the state of the art are
calculations up to the next-to-leading order in renormalization group
improved perturbation theory \cite{bur}.  Potential new physics
contributions enter \eq{p0} through extra contributions to the
$y_i$'s. The long distance physics is non-perturbative and more
difficult to calculate.  It is contained in the hadronic matrix
elements of local four-quark operators $Q_i$:
\begin{eqnarray}
\; \langle Q_i \rangle_I &=& \bra{\lt(\pi \pi\rt)_I } Q_i \ket{K}  \no
\end{eqnarray}
This notation implicitly excludes the strong hadronic phases, which
are factored out of $A_{0,2}$ in \eq{iso}, so that the $\langle Q_i
\rangle_I$'s are real.\footnote{In \cite{fab} the strong phases are 
included in the definition of $\langle Q_i \rangle_I$.} The strong phases stem 
solely from elastic $\pi \pi$ final state rescattering \cite{w}.
The 10 operators involved in \eq{p0} are 
\begin{equation}   
\begin{array}{rcl}
Q_{1} & = & \lt( \overline{s}_{\alpha} u_{\beta}  \rt)_{\rm V-A}
            \lt( \overline{u}_{\beta}  d_{\alpha} \rt)_{\rm V-A}
\, , \\[1ex]
Q_{2} & = & \lt( \overline{s} u \rt)_{\rm V-A}
            \lt( \overline{u} d \rt)_{\rm V-A}
\, , \\[1ex]
Q_{3,5} & = & \lt( \overline{s} d \rt)_{\rm V-A}
   \sum_{q} \lt( \overline{q} q \rt)_{\rm V\mp A}
\, , \\[1ex]
Q_{4,6} & = & \lt( \overline{s}_{\alpha} d_{\beta}  \rt)_{\rm V-A}
   \sum_{q} ( \overline{q}_{\beta}  q_{\alpha} )_{\rm V\mp A}
\, , \\[1ex]
Q_{7,9} & = & \frac{3}{2} \lt( \overline{s} d \rt)_{\rm V-A}
         \sum_{q} e_q \lt( \overline{q} q \rt)_{\rm V\pm A}
\, , \\[1ex]
Q_{8,10} & = & \frac{3}{2} \lt( \overline{s}_{\alpha} 
                                                 d_{\beta} \rt)_{\rm V-A}
     \sum_{q} e_q ( \overline{q}_{\beta}  q_{\alpha})_{\rm V\pm A}
\, . 
\end{array}  
\label{Q1-10}
\end{equation}   
Here $\alpha,\beta$ are colour indices, $e_q$ is the quark electric
charge and $ \lt(\overline{q} q \rt)_{\rm V\pm A}$ is shorthand for
$\gamma_{\nu} (1\pm \gamma_5)$.  Finally isospin breaking from the quark
masses ($m_u\!\neq\! m_d$) is parameterized by
$\Omega$ in \eq{p0}. The dominant contribution to
$\Pi_0$ in \eq{p0} comes from the QCD-penguin operator $Q_6$ generated
by the first diagram of Fig.~\ref{peng}. Likewise $\Pi_2$ is dominated
by the Z-penguins in Fig.~\ref{peng}, which generate
$Q_8$. Graphically one can obtain the operators in \eq{Q1-10} by
contracting the corresponding Feynman diagrams to a point.

\section{Hadronic matrix elements}\label{hama}
Here we focus on the dominant operators $Q_6$ and $Q_8$ in \eq{p0}.
Their hadronic matrix elements are commonly parameterized in terms of 
bag factors $B_6^{(1/2)}$ and $B_8^{(3/2)}$:
\begin{eqnarray}
\; \langle Q_6\rangle_0 \!\!&=&\!\! -4 \sqrt{\frac{3}{2}} 
	\lt( \frac{m_K^2}{m_s\lt(\mu\rt) + m_d \lt(\mu\rt) } \rt)^2 
	\cdot \nn
&& \phantom{-4 \sqrt{\frac{3}{2}} }  \lt( F_K -F_\pi \rt) 
	B_6^{(1/2)} (\mu) \no\\[1ex] 
\; \langle Q_8 \rangle_2 \!\!&=&\!\! \sqrt{3} \, F_\pi \lt[ 
	\lt( \frac{m_K^2}{m_s\lt(\mu\rt) + m_d \lt(\mu\rt) } \rt)^2
	 \,  \rt. \label{b} \\ 
&& \lt. \phantom{\sqrt{3} F } 
	- \frac{1}{6} \lt( m_K^2-m_\pi^2 \rt) \rt] 
	B_8^{(3/2)} (\mu) . \no 
\end{eqnarray}
Here $m_K$, $m_\pi$, $F_K=160\mev$ and $F_\pi=132\mev$ are the masses
and decay constants of Kaon and Pion.  The renormalization scheme and
scale $\mu$ of the quark masses $m_s$ and $m_d$ and the bag factors
$B_6^{(1/2)}$, $B_8^{(3/2)}$ are defined by the scheme and scale
chosen for the operators $Q_6$ and $Q_8$. The vacuum insertion
approximation corresponds to $B_6^{(1/2)}=B_8^{(3/2)}=1$. 
$\mu$ is chosen to be a scale of order $1\gev$, high enough for 
QCD perturbation theory to be trusted and low enough for
non-perturbative methods to work.

Physical observables like $\epsilon^\prime$ do not depend on the
renormalization scheme and scale. Hence ideally the dependence on the
unphysical scale $\mu$ and the renormalization scheme cancels between
the hadronic matrix elements in \eq{b} and the Wilson coefficients
$y_6(\mu)$ and $y_8(\mu)$ multiplying them in \eq{p0}. This
cancellation indeed occurs in those non-perturbative methods, which
use the same degrees of freedom as the perturbative calculation
(quarks and gluons) and allow for a matching calculation between the
short distance physics contained in the $y_i$'s and the long-distance
physics residing in the operator matrix elements in \eq{b}. The only
known method with this feature is lattice gauge theory. In other
methods the remaining dependence on the scheme and scale can be used
to estimate the `theoretical uncertainty' of the calculation. 

I want to stress here that there is no shortcut to solve this problem:
The operator product expansion lumps the physics from all scales
larger than $\mu$ into the Wilson coefficients, while the matrix
elements comprise the dynamics associated with scales smaller than
$\mu$. This enforces heavy mass scales like the top mass to enter the
$y_i$'s rather than the $\langle Q_i\rangle$'s, but it still leaves
the freedom to assign any constant factor either to the coefficients
or to the matrix elements. Changing the scale $\mu$ or the scheme
chosen to calculate the $y_i$'s shuffles such constant factors from
the Wilson coefficients to the matrix elements. The quality of the
scheme and scale cancellations between coefficients and hadronic
matrix elements therefore measures how smoothly the hadronic method
chosen to calculate the $\langle Q_i\rangle$'s merges into
perturbative QCD at the scale $\mu={\cal O} (1\gev)$.  In the
literature one can find attempts to cancel this ambiguity in an ad hoc
way by adding the perturbatively calculated matrix element to the
Wilson coefficients. This formally cancels the scale and scheme
dependence of the latter, but introduces the dependence on the
infrared regulator instead, which is just another unphysical
parameter. 

It is important to note that the dominant $\mu$ dependence of 
$Q_6$ and $Q_8$ in \eq{b} is reproduced by the $\mu$ dependence 
of the quark masses. Hence  $B_6^{(1/2)}$ and $B_8^{(3/2)}$ depend 
only very weakly on $\mu$, so that I omit the reference to $\mu$ 
in the following. 

Next I summarize the three standard methods to calculate the hadronic
parameters $B_i$:\\[0.5ex] 
{\bf Lattice gauge theory} solves non-perturbative QCD on a discrete
spacetime lattice. It controls the scheme and scale dependence of the
Wilson coefficients exactly. At present the $B$-parameters are
calculated in the quenched approximation, i.e.\ without dynamical
fermions. The error caused by this cannot be reliably
estimated. Further present calculations determine $\bra{\pi } Q_i
\ket{K}$ and $\bra{0 } Q_i \ket{K}$ on the lattice and relate the
result to $\bra{\lt(\pi \pi\rt)_I } Q_i \ket{K}$ using lowest order
chiral perturbation theory, thereby introducing additional model
dependence.  Finally not all systematic errors associated with the
discretization of QCD are fully understood yet. Recent lattice results
for $B_8^{(3/2)}$ in the $\ov{\rm MS}$ NDR scheme are
\begin{eqnarray}
B_8^{(3/2)} \!\!\!\!&=& \!\!\!\left\{ 
\begin{array}{l@{~}l}
0.77(4)(4) & \cite{kgs} \\ 
0.81(3)(3) & \cite{gbs} \\
0.82(2)    & \cite{ape} \\
1.03(3)    & \mbox{(non-pert.\ matching) \cite{ape}.} 
\end{array}\rt. \label{b8}
\end{eqnarray}
The calculation of $I=0$ amplitudes is more difficult. A recent
calculation of $B_6^{(1/2)}$ \cite{blum} using the new method of
domain wall fermions has found a negative $B_6^{(1/2)}$, which sharply
contradicts the result obtained by other methods and is hardly
compatible with experiment even in the presence of new physics
\cite{bf}.  \\[0.5ex] The {\bf $\bf 1/N_c$ expansion} is a rigorous
QCD-based method, too. The expansion parameter is the inverse number
of colours, $1/N_c=1/3$. The leading order corresponding to
$N_c=\infty$ consists of all planar QCD Feynman diagrams
\cite{tH}. These diagrams correspond to tree-level diagrams in an
effective meson theory, which is based on a chiral lagrangian
$\chi{\cal L}$ \cite{nc}. Likewise $1/N_c$ corrections correspond to
one-loop diagrams in the meson theory. The loop integrals are
calculated with an explicit cutoff $\Lambda$, which separates the low
energy hadronic region calculated from $\chi{\cal L}$ from the high
energy perturbative region of QCD. One can show that $\Lambda$
appearing in the matrix elements and $\mu$ contained in the Wilson
coefficients are proportional to each other, usually they are taken to
be equal. Present calculations yield a quadratic dependence of the
matrix elements on $\Lambda$, which is expected to turn into the
correct logarithmic behaviour of the coefficients once vector mesons
are included in the calculation.  The scale and scheme dependence of
the $y_i$'s are exactly cancelled for $N_c=\infty$ and are
qualitatively under control at order $1/N_c$. A complete control of
the scheme dependence at this order is expected to be possible
\cite{ba}. In the large-$N_c$ limit one has
$B_6^{(1/2)}=B_8^{(3/2)}=1$.  Recently $p^0/N_c$ and $p^2$ corrections
to $B_6^{(1/2)}$ and $B_8^{(3/2)}$ in the combined $1/N_c$ and chiral
expansion have been calculated \cite{do}. The predicted ranges for the
parameters are
\begin{eqnarray}
\; 0.42 \leq B_8^{(3/2)} \leq 0.64, && \!\!\!\!\! \!\! 
0.72 \leq B_6^{(1/2)} \leq 1.30 \label{resnc}
\end{eqnarray}
showing that the corrections to the large $N_c$ limit are indeed
reasonably small, as expected in \cite{nc,bjl,rev1}. The situation is
different in the case of $B_1^{(1/2)}$ and $B_2^{(1/2)}$, which
receive anomalously large $1/N_c$ corrections in qualitative
phenomenological agreement with the large observed value of $\real
A_0$.  It is common practice to extract the $B$-parameters of the
subdominant operators in \eq{p0} from the experimental values of
$\real A_0$ and $\real A_2$ \cite{bjl}.
\\[0.5ex] 
The {\bf chiral quark model} \cite{chqm,rev2} calculates all
$B$-parameters in terms of three model parameters, which are determined
from $\real A_0$ and $\real A_2$.  It includes chiral corrections up
to ${\cal O} (p^4)$. The chiral quark model shows a qualitative
control of the scale dependence. As in any other model, it is
difficult to judge systematic errors of the chiral quark model. My
presentation of this model is brief here, because it is covered in
some detail in Fabbrichesi's talk \cite{fab}.  In the $\ov{\rm MS}$
HV scheme the predictions for the $B$-parameters read:
\begin{eqnarray}
\; 0.75 \leq B_8^{(3/2)} \leq 0.94 && \; 
1.1 \leq B_6^{(1/2)} \leq 1.9. \label{reschqm}
\end{eqnarray}

\section{Phenomenology}
The Standard Model prediction for $\epsilon^\prime/\epsilon$ can be
summarized in the handy approximate formula \cite{bbg}:
\begin{eqnarray}
\frac{\epsilon^\prime}{\epsilon}\! &=& \! 21 \cdot 10^{-4} \, 
 \frac{\imag \lambda_t}{1.7\cdot 10^{-4}} \, 
	\lt[ \frac{100\mev}{m_s(2\gev)} \rt]^2 \cdot \label{esm}\\
&& \!
	\lt[  B_6^{(1/2)} \frac{1-\Omega}{0.8} 
	-0.5 \,  B_8^{(3/2)} \rt] \frac{\Lambda_{\ov{\rm MS}}}{340\mev}   
\no.
\end{eqnarray}
Here $\Lambda_{\ov{\rm MS}}$ is the fundamental scale parameter of QCD
\cite{bbdm}. $\Lambda_{\ov{\rm MS}} \!=\! 340\mev$ corresponds to
$\alpha_s(M_Z)\!=\!0.119$.  $\imag \lambda_t$ must be determined from a
standard ana\-lysis of the unitarity triangle using $\epsilon$,
$|V_{cb}|$, $|V_{ub}|$ and the mass differences $\Delta m_{d,s}$ of
$B_{d,s}$ mesons. The constraint from $\e$ in \eq{expeps} on $\imag
\lambda_t$ reads
\begin{eqnarray}
\; 6.0 \cdot 10^{-8} \!\!&=&\!\!  \hat{B}_K \, \imag \lambda_t \, \lt[ 
	\real \lambda_c \, \lt[ \eta_1 S_{cc} - \eta_3 S_{ct} \rt] \rt.  \nn
&& \phantom{ B_K \, \imag \lambda_t \,} 	
	\lt. - \real \lambda_t \, \eta_2  
	  S_{tt}  \rt] . \label{eps}
\end{eqnarray}
Here the result of he box diagram of Fig.~\ref{box} is contained in
$S_{cc}= x_c$, $S_{ct}=x_c (0.6-\log x_c)$ and $S_{tt}=2.5$ with
$x_c=m_c^2/M_W^2$ containing the charm quark mass $m_c\!\! \approx \!\! 1.3\gev$
in the $\ov{\rm MS}$ scheme and the W-boson mass. The well-measured top
quark mass has entered the numerical constants 0.6 and 2.5
here. $\eta_1=1.4\pm0.2$, $\eta_2=0.57\pm0.01$ and $\eta_3=0.47 \pm
0.04$ are QCD correction factors \cite{etas}. $\hat{B}_K$ in \eq{eps}
parameterizes the hadronic matrix element of $\Delta S=2$ operator
generated by the box diagram in Fig.~\ref{box}. The analyses of the 
unitarity triangle \cite{etas,bbg} yield:
\begin{eqnarray}
\; 1.0\cdot 10^{-4} \leq \imag \lambda_t \leq 1.7 \cdot 10^{-4}. \label{imla}
\end{eqnarray}
The choice of a certain method to calculate the hadronic parameters
affects both the phenomenology of $\e$ in \eq{eps} and of
$\e^\prime/\e$ in \eq{esm}.  Thus it correlates $\imag\lambda_t$ in
\eq{esm} with $B_6^{(1/2)}$ and $B_8^{(3/2)}$. For example the
chiral quark model \cite{chqm} predicts a higher value for $\hat{B}_K$
than lattice or $1/N_c$ calculations. Therefore one extracts a smaller
value for $\imag \lambda_t$ than in \eq{imla}, where the results of the
latter two methods have been used. It must be stressed, however, that
the experimental upper limit in \eq{imla} stems solely from the upper
bounds on $|V_{ub}|$ and $|V_{cb}|$. $\epsilon_K$ has no influence on
the upper limit of $\imag \lambda_t$, which is obtained by setting 
$\sin \gamma =1$ in \eq{imlat}. 

The strong dependence of $\e^\prime/\e$ in \eq{esm} on $m_s$ is
another big source of uncertainty in the prediction of $\e^\prime/\e$.
QCD sum rule calculations favour the range \cite{sr}
\begin{eqnarray}
\;m_s(2 \gev)=124\pm22\mev. \label{srms}
\end{eqnarray}
There are now three quenched lattice calculations which control all
systematic errors \cite{latt}. They are nicely consistent with each
other and predict
\begin{eqnarray}
\;m_s(2 \gev)=103\pm 10\mev . \label{lattms}
\end{eqnarray}
Further the CP-PACS collaboration \cite{pacs} has reported the result
of an unquenched calculation yielding $m_s(2 \gev)=84\pm 7\mev$.  The
determination of the strange quark mass from Cabibbo suppressed $\tau$
decays has been recently clarified by Pich and Prades, who extract
\begin{eqnarray}
\;m_s(2 \gev)=114\pm23\mev \label{ppms}
\end{eqnarray}
from ALEPH data \cite{pp}.

Finally $\Omega=0.25\pm0.08$ \cite{iso} completes
the list of the input parameters of $\e^\prime/\e$. 
From \eq{esm} one notes that the $\Delta I=1/2$ and $\Delta I=3/2$
contributions tend to cancel each other, which increases the
uncertainty of the prediction. $B_6^{(1/2)}$ and $B_8^{(3/2)}$ 
enter \eq{esm} essentially in the combination 
$2\, B_6^{(1/2)}-B_8^{(3/2)}$. It is now easy to see from \eq{esm}
that it is difficult to fit the high measured value in \eq{exp} with
the $B$-parameters in \eq{b8} and \eq{resnc}. A recent detailed
analysis \cite{bbg} found  
\begin{eqnarray}
\; \frac{\e^\prime}{\e} &=& \lt( 7.7 \epm{6.0}{3.5} \rt) \cdot 10^{-4},
 \label{muc}
\end{eqnarray}
substantially below the experimental result. On the other hand it is
possible to reach the value in \eq{exp}, if simultaneously 
 $2\, B_6^{(1/2)}-B_8^{(3/2)}$ is large and $m_s$ is small \cite{bbg,kns}.
In \cite{kns} an upper bound on $m_s$ has been derived from the
requirement that $\e^\prime/\e \geq 20 \cdot 10^{-4}$ corresponding to
the 2$\sigma$ bound of the KTeV measurement \cite{ktev}. If 
$2\, B_6^{(1/2)}-B_8^{(3/2)}\leq 2.0$ suggested by \eq{b8} and
\eq{resnc}, then 
\begin{eqnarray} 
\; m_s \lt( 2\gev \rt) \leq 110 \mev. \label{ms}
\end{eqnarray} 
This is in agreement with \eq{lattms} and \eq{ppms}, but not with all
sum rule results representated by the range in \eq{srms}. As discussed 
at this conference, especially values for  $m_s \lt( 2\gev \rt)$ below
$100\mev$ are in serious conflict with QCD sum rule calculations. Such
low values would imply that the breakdown of the perturbative
calculations of the relevant spectral functions sets in at a much
higher scale than commonly expected.  

The chiral quark model allows for a larger range for $2\,
B_6^{(1/2)}-B_8^{(3/2)}$ in \eq{reschqm}. It predicts \cite{rev2}
\begin{eqnarray}
\; \frac{\e^\prime}{\e} &=& \lt( 17 \epm{14}{10}\rt)\cdot 10^{-4},
 \label{trs}
\end{eqnarray}
in better agreement with the data in \eq{exp}. The difference between
\eq{trs} and \eq{muc} stems not only from the different ranges for the
$B$-parameters in \eq{reschqm} and \eq{resnc} but also from a
different treatment of the final state phases: None of the methods
described in sect.~\ref{hama} predicts the strong phases
$\delta_{0,2}$ correctly. Lattice calculations cannot do this, because
they are performed with Euclidean time. The other two methods find
$\delta_{0}$ much smaller than the experimental value
$\delta_{0}^{exp}=37^\circ$. This leads to some ambiguity in the
prediction for $\e^\prime/\e$, because one can either identify the
magnitude or the real part of the calculated $\langle Q_6 \rangle_0
\exp (i \delta_0^{calc})$ with the desired true $\langle Q_6 \rangle_0
\exp (i \delta_0^{exp})$. The former identification has been chosen in
\cite{nc,bjl,rev1} with the result quoted in \eq{muc}, while the
Trieste group \cite{fab,chqm,rev2} has used the real part in their
prediction of \eq{trs}. One can implement the Trieste method into the
$1/N_c$ result by rescaling the range for $B_6^{(1/2)}$ in \eq{resnc}
by a factor of $1/\cos(\delta_{0}^{exp})=1.25$ to $0.9 \leq
B_6^{(1/2)} \leq 1.6$. This increases the range for $\e^\prime/\e$ in
\eq{muc} and relaxes the upper bound on $m_s (2\gev)$ in \eq{ms} to
125 \mev.  Hence a part of the discrepancy between \eq{muc} and
\eq{trs} stems from the treatment of the strong phases and has nothing
to do with the different treatment of the strong dynamics.

Do we need $m_s$ at all to predict $\e^\prime/\e$? The strange quark
mass enters the hadronic matrix elements in \eq{b}, because they are
normalized to their value in the vacuum insertion approximation.  In
the distant future lattice calculations might directly compute the
matrix elements in \eq{b} proportional to $B_i/m_s^2$ rather than
$m_s$ and the $B_i$'s separately. Yet it should be stressed that
present lattice results summarized in \eq{b8} stem from calculations
of the B-parameter $B_8^{(3/2)}$ rather than the full matrix element.
The importance of $m_s$ for the prediction of $\e^\prime/\e$ in
lattice calculations has been stressed in \cite{m}. Also the $1/N_c$
expansion naturally introduces the factor $1/m_s^2$ into \eq{b}: In
the large-$N_c$ limit the matrix elements of $Q_6$ and $Q_8$ reduce to
those of density currents $\ov{s}q_{S+P}=\ov{s} (1+\gamma_5) q$ and
$\ov{q}d_{S-P}$. They are related to vector currents by the equation
of motion, which introduces $m_s$ into the result. The $B$-parameters
obtained in the $1/N_c$ expansion are independent of $m_s$. Other
methods express matrix elements of (density)$\times$(density)
operators in terms of the quark condensate, which is related to $m_s$
and $m_K$ by the PCAC relation. In these methods the
parametrization in \eq{b} may be unnatural and the thereby defined
$B$-parameters can exhibit a sizeable dependence on $m_s$. E.g.\ in the
chiral quark model $B_6^{(1/2)}$ is proportional to $m_s$.

\begin{figure*} 
\centerline{\epsfxsize=0.9\textwidth
		\epsffile{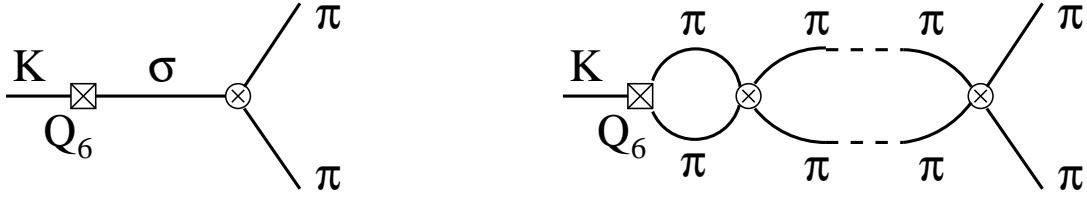}}
\caption{The $\sigma=f_0(400-1200)$ resonance enhances 
$\bra{\lt(\pi\pi\rt)_0 } Q_i \ket{K} \propto B_6^{(1/2)} $. It
corresponds to  resonant $\pi\pi$ rescattering in the $I=0$ channel as
depicted by the right diagram. The crossed square denotes the weak
interaction mediated by $Q_6$ and the crossed circles represent strong
interaction vertices.}\label{sig}
\end{figure*}

The second possibility to reproduce \eq{exp} is to consider higher
values for $B_6^{(1/2)}$, as suggested by the chiral quark model in
\eq{reschqm}. An important feature of the chiral quark model is the
proliferation of the $\Delta I=1/2$ enhancement present in $\real A_0$
(cf.\ \eq{d1/2}) into $B_6^{(1/2)}$. The three parameters of the
chiral quark model are determined from the $\Delta I=1/2$ rule which
thereby feeds into $B_6^{(1/2)}$. The enhancement of $\real A_0$
originates from penguin contractions of the operators $Q_1$ and $Q_2$
\cite{nc,bbg}, which involve the same spin and isospin quantum numbers
as $Q_6$.  The fact that the chiral quark model can reproduce \eq{exp}
more easily could indicate that a $\Delta I=1/2$ enhancement is at
work in $B_6^{(1/2)}$ as well \cite{bk99}.  This hypothesis has also
been recently stressed by the Dortmund group \cite{do2}, which found
indication of a further enhancement of $B_6^{(1/2)}$ from a higher
order term (${\cal O} (p^2/N_c)$) in the $1/N_c$ expansion. It is
worthwile to notice that the $1/N_c$ expansion fails to simultaneously
reproduce both $\real A_0$ and $\real A_2$ \cite{do3}.  The second
reason for the large value of $\e^\prime/\e$ in \eq{trs} is the
special treatment of the final state phases as described after
\eq{trs}, which is controversial \cite{bbg}. It is highly desirable to
gain a full dynamical understanding of both the $\Delta I=1/2$
enhancement in $\real A_0$ and possibly in $B_6^{(1/2)}$ and the large
final state interaction phase $\delta_0=37^\circ$ causing the
discussed ambiguity in $\e^\prime/\e$.  To this end in \cite{kns,mls}
a resonant enhancement of $\langle Q_6 \rangle_0$ caused by the
$\sigma=f_0(400-1200)$ \cite{pdg} resonance has been
discussed. $\sigma$ is a broad S-wave $I=0$ resonance almost
degenerate in mass with the Kaon.  In theories based on a chiral
lagrangian $\sigma$ corresponds to resonant $\pi\pi$ rescattering in
the $I=0$ channel, an all-order effect depicted in Fig.~\ref{sig}.
Unfortunatley the progressing lattice calculations will not help to
understand such a resonant enhancement of $B_6^{(1/2)}$, because they
will determine $\bra{\pi } Q_i \ket{K}$ rather than $\bra{\lt(\pi
\pi\rt)_0 } Q_i \ket{K}$! Even if reliable lattice results for
$\bra{\pi } Q_i \ket{K}$ are present, there will still be a sizeable
model dependence in the prediction of $B_6^{(1/2)}\propto \bra{\lt(\pi
\pi\rt)_0 } Q_i \ket{K}$.

Finally another source of theoretical uncertainties has been recently 
suggested \cite{gv}: Isospin violations in $\langle Q_6 \rangle$ 
resulting from $m_u\!\neq\!m_d$ can substantially enlarge the range 
$0.15 \leq \Omega \leq 0.35$ used in \cite{rev2,bbg}. 

\section{New physics}
Since $\e^\prime/\e$ is a short distance dominated process, it is
sensitive to new physics, which can modify the Wilson coefficients
$y_i$. The $y_i$'s are generated by Feynman diagrams like those in
Fig.~\ref{peng} at some high scale $\mu$ of the order of $M_W$ or of
the mass of some new particle entering the loop diagrams. The
renormalization group evolution down to the scale $\mu\! \approx \!
1\gev$, at which the matrix elements in \eq{p0} are calculated, mixes
the coefficients $y_i$.  Now $y_6(\mu\!\! \approx \!\! 1\gev)$ is
mainly an admixture of the Wilson coefficient of $Q_2$, which
is generated by tree-level $W$-exchange and is therefore hardly
affected by new physics. On the other hand $y_8(\mu\!  \approx \!
1\gev) $ is merely loop-induced and stems mainly from the
$\ov{s}dZ$-vertex (see Fig.~\ref{peng}). It is therefore sensitive to
new physics.  If new contributions to $y_8$ have the opposite sign of
the Standard Model contribution, $\e^\prime/\e$ will be
enhanced. Further the chromomagnetic operator
\begin{eqnarray} 
\; Q_{11} & = & \frac{g_s}{16 \pi^2} m_s \ov{s} \sigma^{\mu \nu} T^a
                (1-\gamma_5) d \, G_{\mu \nu}^a \label{chr}.
\end{eqnarray}
can play a role. In the Standard Model its coefficient equals
$y_{11}(\mu \! \approx \!1\gev ) \! \approx \! -0.19$ and the
impact on $\e^\prime/\e$ is negligible \cite{chqm}, so that $y_{11}
\langle Q_{11} \rangle_0$ has been omitted in \eq{p0}. A
model-independent discussion of new contributions to $y_8$ and
$y_{11}$ has been performed in \cite{kns}. An enhancement of the
Standard Model result for $y_{11}$ by a factor of order $500$ is
necessary for a sizeable impact on $\e^\prime/\e$.

The high experimental value in \eq{exp} has stimulated new theoretical
work on possible new physics contributions to $\e^\prime/\e$, mainly
in supersymmetric theories. Supersymmetry (SUSY) does not help to
understand the puzzles of flavour physics. Moreover, the minimal
supersymmetric standard model with the most general soft SUSY-breaking
mechanism leads to unacceptably large flavour-changing neutral
transitions. Hence additional assumptions on the SUSY-breaking terms
are necessary, most commonly flavour-blindness of these terms
(``universality'') at some high scale. Yet in these scenarios the
impact of SUSY on $\e^\prime/\e$ is small, and in most of the
parameter space $\e^\prime/\e$ is depleted rather than enhanced
\cite{gi}.  Alternatively one can relax the universality assumption
and allow for arbitrary flavour off-diagonal entries in the squark
mass matrix. Here one proceeds phenomenologically, encounters all
experimental constraints from other flavour-changing processes
\cite{ggms} and finally estimates the maximal impact on $\e^\prime/\e$
or other processes of interest.  In these generic SUSY models,
however, the situation for $\e^\prime/\e$ is different than in
scenarios with flavour universality: It is possible to have sizeable
contributions to the imaginary parts of the $\ov{s}dZ$-vertex
affecting $y_8$ \cite{ci} and the chromomagnetic
$\ov{s}d$-gluon-vertex modifying $y_{11}$ \cite{bfm,mm} without
violating the stringent bounds from Kaon mixing and other
flavour-changing processes. Then $\e^\prime/\e$ can even be dominated
by new physics. In \cite{mm} an approximate flavour symmetry
controlling the Yukawa matrix and the SUSY A-matrix has been
postulated. Then the small weak $s\rightarrow d$ transition
proportional to $\imag \lambda_t={\cal O}(\lambda^5)$ in \eq{eps} is
replaced by a strong transition of order $\lambda$ producing the
necessary enhancement factor. A recent detailed analysis \cite{bcirs}
has found an enhancement of $\e^\prime/\e$ through $y_{11}$ more
likely than through $y_8$. Finally in addition a mass splitting
between up and down squarks also generates new $\Delta I=3/2$
contributions via box diagrams which likewise influence $\e^\prime/\e$
\cite{nk}.

\section{Conclusions and outlook}
It is difficult, but possible to accomodate $\e^\prime/\e\simeq 2\cdot
10^{-3} $ in the Standard Model. I have discussed the following
mechanisms to explain the large measured value in \eq{exp}:
\begin{itemize}
\item[1)] Small $m_s$ as indicated by recent lattice results 
	  \cite{latt,pacs}.
\item[2)] $\Delta I=1/2$ enhancement of $B_6^{(1/2)}$ \cite{kns,bk99,do2}. 
\item[3)] Larger isospin breaking characterized by the parameter 
	  $\Omega$ \cite{gv}.   
\item[4)] Decrease of the imaginary part of the $\ov{s}dZ$-vertex 
 	  (and thereby of $y_8$) by new physics \cite{ci,bcirs}. 
\item[5)] ${\cal O} (500)$ enhancement of the chromomagnetic
	  $\ov{s}d$-gluon-vertex (and thereby of $y_{11}$) by new physics
	  \cite{bfm,mm,bcirs}.
\end{itemize}
In the future one can expect improvements in the prediction of
$\e^\prime/\e$ from a better knowledge of $m_s$ from $\tau$-decays
\cite{pp} and from a better determination of $\imag \lambda_t$, once
the dedicated B- and K-experiments \cite{babar,fnal,k+,k0} will give
us a clear picture of the unitarity triangle. On the other hand I do
not expect much progress in $B_6^{(1/2)}$ from lattice calculations,
because these calculations will not determine $\bra{\lt(\pi \pi\rt)_0
} Q_i \ket{K}$ directly. Hence the question of new physics in
$\e^\prime/\e$ will stay inconclusive and the corresponding new
contributions will more likely be revealed in the measurements
\cite{k+,k0} of rare Kaon decays \cite{ci,bcirs}.
 
\section*{Acknowledgements}
I thank Matthias Neubert for inviting me to this conference and
Stephane Narison and his team for creating the pleasant and stimulating
atmosphere. I have enjoyed a lot of fruitful discussions with many
colleagues. I am grateful to Andrzej Buras for proofreading the
manuscript.

~

{\sc Comment ({\bf M.~Knecht, CPT/CNRS Marseille)}}: {\it I have a
comment to A.~Pich's comment concerning the dependence of
$\e^\prime/\e$ on the size of the quark condensate. If one does the
things correctly within the framework of generalized $\chi$PT, one
finds that the prediction for $\e^\prime/\e$ can easily be
\underline{increased} by a factor of 3, taking all other parameters
fixed, i.e.\ $m_s(2 \gev)\sim 150\mev$, if the condensate
was \underline{smaller} by a factor of 10.}

\end{document}